\documentclass{elsarticle}
\usepackage{latexsym}
\usepackage{float}
\usepackage[latin2]{inputenc}
\usepackage{graphicx}
\usepackage{ae}
\usepackage{aecompl}
\usepackage{setspace}
\usepackage{amsmath}
\usepackage{amssymb}
\usepackage{amsthm}
\usepackage[hidelinks]{hyperref}

\makeatletter
\def\ps@pprintTitle{%
	\let\@oddhead\@empty
	\let\@evenhead\@empty
	\def\@oddfoot{\centerline{\thepage}}%
	\let\@evenfoot\@oddfoot}
\makeatother

\begin{document}

\title{PDB\_Amyloid: The Extended Live Amyloid Structure List from the PDB}
	
\author[p]{Kristóf Takács}
\ead{takacs@pitgroup.org}
\author[p]{Bálint Varga}
\ead{balorkany@pitgroup.org}
\author[p,u]{Vince Grolmusz\corref{cor1}}
\ead{grolmusz@pitgroup.org}
\cortext[cor1]{Corresponding author}
\address[p]{PIT Bioinformatics Group, Eötvös University, H-1117 Budapest, Hungary}
\address[u]{Uratim Ltd., H-1118 Budapest, Hungary}

\date{}

\begin{abstract}
	The Protein Data Bank (PDB) contains more than 135 000 entries today. From these, relatively few amyloid structures can be identified, since amyloids  are insoluble in water. Therefore, mostly solid state NMR-recorded amyloid structures are deposited in the PDB. Based on the geometric analysis of these deposited structures we have prepared an automatically updated webserver, which generates the list of the deposited amyloid structures, and, additionally, those globular protein entries, which have amyloid-like substructures of a given size and characteristics. We have found that applying only the properly chosen geometric conditions, it is possible to identify the deposited amyloid structures, and a number of globular proteins with amyloid-like substructures. We have analyzed these globular proteins and have found proofs in the literature that many of them are known to form amyloids more easily than many other globular proteins. Our results relate to the method of (Stankovic, I. et al. (2017): Construction of Amyloid PDB Files Database. Transactions on Internet Research. 13 (1): 47-51), who have applied a hybrid textual-search and geometric approach for finding amyloids in the PDB.
	If one intends to identify a subset of the PDB for some applications, the identification algorithm needs to be re-run periodically, since in 2017, on average, every day 30 new entries were deposited in the data bank. Our webserver is updated regularly and automatically, and the identified amyloid- and partial amyloid structures can be viewed or their list can be downloaded from the site \url{https://pitgroup.org/amyloid}. 
\end{abstract}

\maketitle
	
\section*{Introduction} 

The Protein Data Bank (PDB) is a permanently developing public resource of spatial structures of proteins and nucleic acids \cite{PDB-base}. Today the database contains more than 135,000 structures. The geometric properties of these molecules can be analyzed by bioinformatical tools, and one may infer significant new relations in these very complex macromolecular structures through these analyses \cite{Ivan2014,Ivan2010,Ivan2009,Ivan2007,Ivan2010a,pdbmend,RS-PDB-1,Ordoeg2008}. 

In the present contribution, we are interested in the amyloid structures in the Protein Data Bank. Amyloids are misfolded protein aggregates, which are present in numerous structures in biology, including 

\begin{itemize}
\item the cellular surface of several microorganisms \cite{Gebbink2005,Blanco2012}, having a role in host-pathogen interaction;
\item the silkmoth chorion and some fish choria, forming protective films \cite{Iconomidou2008};
\item the immune system of certain insects, helping the encapsulation of the intruders \cite{Falabella2012};
\item in the healthy human pituitary secretory granules, for storing peptide hormones \cite{Maji2009};
\item in human amyloidoses and several neurodegenerative diseases \cite{Soto2006}.

Amyloid structures sometimes show prion-like infective properties \cite{Soto2006,Caughey2009,Aguzzi2009}. Cerebral $\beta$-amyloid plaques have been considered biomarkers of the Alzheimer's disease for a long time \cite{Alzheimer1907,Prescott2014,Ma2002,Zheng2007}; but more recently, their validity is challenged by several authors \cite{Pimplikar2009,Holmes2008,Hyman2011}.
\end{itemize}

Amyloid formation mechanisms are reviewed in \cite{Eisenberg2012,Richardson2002}. Amyloid fibers are formed from parallel $\beta$-sheets, with hydrogen bonds between the parallel strands. It is widely accepted that amyloid formation requires the presence of a nucleus or a seed of amyloid-forming segments with exposed edges of the $\beta$-sheet structures \cite{Nelson2005,Lee2011,Eisenberg2012,Richardson2002}. 

Since amyloid fibers are insoluble in water until the very recent years there were no high-resolution structures deposited in the RCSB Protein Data Bank (PDB) \cite{PDB-base}. Today one can find several dozens of atomic resolution amyloid structures in the PDB, and this dataset opened up the possibility of the analysis and the data mining of the properties of these misaligned proteins, using their high-resolution spatial structure. 

The first step in this direction is the identification of the amyloid structures in the PDB. 

Amyloid- and amyloid-precursor molecules were collected and predicted using protein-sequencing data in numerous articles (e.g.,  in the AMYPdb resource \cite{LeBechec2008}, or in \cite{Tartaglia2008}). We are interested in the analysis of the {\em spatial} protein structures for finding amyloid and amyloid-precursor molecules, rather than the analysis of residue-sequence properties of proteins of unknown three-dimensional structures.

In the remarkable work \cite{Stankovic2017} the authors screened the Protein Data Bank for amyloid structures, by applying the following procedure:

\begin{itemize}
	
\item[(i)] By textual search, those PDB entries were selected, which contain the words ``amyloid'' or any of other 38 words, describing amyloid precursors;

\item[(ii)]	Helical structures, identified by torsion angles, were thrown out;

\item[(iii)] Parallel, near-linear fragments of length at least 4 residues were identified; structures without these fragments were also thrown away.
\end{itemize}

In the present work, we prepare an automatically updated list of amyloid and potentially amyloidogenic structures from the PDB, with applying only the geometric properties of $\beta$-sheets; consequently, we do not use any textual search, referring to the annotations of the PDB entries.

By this choice, we intend to identify not only the aggregated amyloid entries and known precursors, but also those globular proteins, which contain small, locally amyloid-like substructures. We assume that these globular proteins may also be amyloidogenic ones, i.e., they can more easily turn into amyloid fibers than globular proteins without these structural elements.
 
Since the Protein Data Bank grows very quickly -- in 2017, every day, on the average, 30 new structures were deposited -- we need to construct an automatically updated web server, which periodically examines the new PDB entries and includes the newly deposited amyloid and potentially amyloidogenic structures. Consequently, our list does not give a snapshot of the amyloid structures in the PDB at a given time as other efforts but rather presents a {\em live list} of these structures. Our online resource is available at \url{https://pitgroup.org/amyloid/}.

\section*{Discussion and Results}

\subsection*{Amyloid structures}

We have found that our list at \url{https://pitgroup.org/amyloid/} contains all amyloid structures with at least two polypeptide chains, which are listed in \cite{Stankovic2017}. For example, the classical amyloid structures of 2KIB, 2N0A, 5KO0, 2LBU, 2LMN are all present in the list. 

\subsection*{Possibly Amyloidogenic Structures}

Here we review some non-amyloid proteins which were found by our screening algorithm, and which are listed at \url{https://pitgroup.org/amyloid/}. We also give literary evidence showing links to the amyloid formation of these molecules. These findings witness the power of our algorithm, but clearly, we cannot review here the more than 500 structures, presented on the webpage \url{https://pitgroup.org/amyloid/}.

\begin{itemize}
	\item[1HCN:] Human chorionic gonadotropin (hCG), Fig. 1, panel A. It is a placenta-produced human hormone, applied in numerous pregnancy tests and in legal and illegal drug products, including physical performance enhancing and weight-loss preparations. It is reported to increase $\beta$-amyloid levels in rats \cite{Berry2008} and to increase $\beta$-cleavage of an amyloid-precursor protein \cite{Saberi2013}. Protein hCG also has a role in amyloid $\beta$ precursor protein expression and modulation in human cells \cite{Porayette2007}, and in protein folding regulation in endoplasmic reticulum \cite{Ruddon1996}. We believe that these roles of hCG are closely related to particular geometric properties of its parallel $\beta$-sheets.
	
	\item[1BSF:] Thymidylate synthase A (TS) from {\em Bacillus subtilis}. Thymidylate synthase has an important role in DNA synthesis, whose aggregational properties were studied for a long time \cite{Agarwalla1996}. The human TS is a primary target of cancer chemotherapy, most importantly by 5-fluorouracil, a strong-binding TS inhibitor, applied widely in colon-, esophageal-, stomach-, pancreatic-, breast- and cervical cancers. On Fig. 1., panel B, it is clearly visible that the parallel $beta$-sheets are hidden in the dimeric structure. As it is shown in \cite{Genovese2010}, TS also has a monomeric form with distinct function, and the dimeric and the monomeric forms have an equilibrium in humans. Therefore, the hidden $\beta$-sheets in the monomeric form may become accessible and may play a role in aggregation processes.
	
	\item[3FJ5:] Tyrosine kinase c-Src (Fig. 1, panel C) has a role in MAP kinase pathway, and in the development of breast cancers in animals and humans \cite{Guy1994,Biscardi2000}. It is shown that the SH3 domain of this protein aggregates to form amyloid fibrils at mild acid pHs in \cite{Bacarizo2014}. Sources \cite{Dhawan2012,Dhawan2012a} suggest that amyloid associated microgliosis is strengthened by tyrosine kinase c-Src activity. It is also noted that MAP kinase signaling cascade dysfunction specific to Alzheimer's disease in fibroblasts \cite{Zhao2002}. 
	
	\item[2OCT:] Stefin B (Cystatin B) tetramer (Fig. 1, panel D) is an intracellular thiol protease inhibitor. It is known to form amyloid fibrils {\em in vitro} \cite{Zerovnik2002},  its role in amyloidogenesis is detailed in \cite{Kokalj2007} and \cite{Skerget2010}.

\end{itemize}

\section*{Materials and Methods}

Here we describe the selection method, which generates the Extended Amyloid List at \url{https://pitgroup.org/amyloid/}. 

In contrast with \cite{Stankovic2017}, we did not make any selection through textual search in the annotation fields of the PDB files. Instead of that, we have attempted to collect the {\em minimal set of geometric rules}, which already return the amyloids found in \cite{Stankovic2017}, plus novel, globular proteins with possibly amyloidogenic substructures.

The following rules are applied:

\begin{itemize}

\item[(i)] For finding parallel $\beta$-sheets: The authors of \cite{Stankovic2017} selected parallel chain segments by requiring the distance-difference between the closest $C_\alpha$ atoms of the fragment to be less than 1.5 \AA. Instead of this condition, we have applied a bound to the standard deviation $\sigma$ between the closest $C_\alpha$ atoms of the fragment to be less than 1.5 \AA. We think that this approach is more tolerant to singular, random errors in the structure, while it is strict enough for characterizing  the parallel polypeptide chains in the amyloid structures. More technically, our condition can be re-phrased as follows: let us consider two separate chains of the structures, $A$ and $B$, both identified as $\beta$-sheets. Next, we compute the array $C(A)_{dist}$, which contains the minimum distance for every $C_{\alpha}$ atom of chain $A$ from the closest $C_{\alpha}$ that is located in chain $B$. Next, we identify the maximal sub-chains $F$ of $A$, satisfying $\sigma_(C(F)_{dist}) \leq 1.5$, while every distance in vector $C(F)_{dist}$ are required to be between 2 and 15 \AA. 

\item[(ii)] Excluding structures with large curvature: The authors of \cite{Stankovic2017} excluded helical structures from consideration. We apply a locally verifiable angular condition for the fragments $F$ as follows: Fragment $F$, which satisfies the conditions in (i), needs also to satisfy the condition that the angles of each three consecutive $C_{\alpha}$ atoms, averaged for the fragment $F$, need to be between  $110 ^{\circ}$ and $180 ^{\circ}$. In other words, these angles, on the average, should be obtuse angles between  $110 ^{\circ}$ and $180 ^{\circ}$.
 
\item[(iii)] Condition for the minimum length of parallel fragments: $len(F) \geq \frac{len(chain_A)}{7}$ where $F$ denotes the same as in $(i)$, and $len(X)$ denotes the length of chain $X$, measured in residues.

\end{itemize}

The specific parameters for the conditions above were selected for including all multi-chain amyloid structures that were also found in \cite{Stankovic2017}. We do not aim to find amyloid-like structures containing only one single polypeptide chain since the amyloid structures contain a large number of approximately parallel fibers, each consisting of different chains. While the Protein Data Bank contains partial amyloid structures with one single chain (e.g., 1HZ3), these structures will not be listed within our results, since they lack the characteristic property of nearly parallel, distinct polypeptide chains.  The search for {\em distinct} parallel chains is useful for disallowing single chains with long parallel subsections, for example, beta-barrel structures, like the bacterial porin structure 4RLC. Since amyloid aggregates always consist of a large number of distinct, approximately parallel polypeptide chains, our condition is not restrictive.

\begin{figure}[H]
	\begin{center}
		\includegraphics[width=12cm]{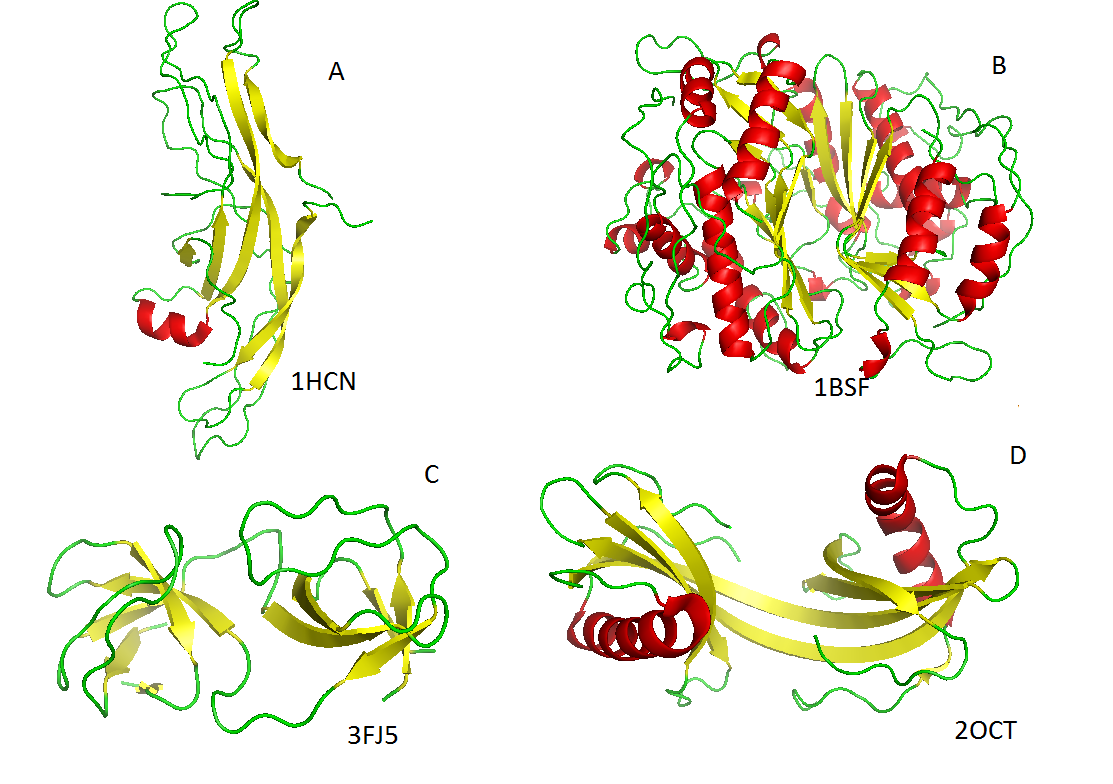}
		\caption{Protein structures 1HCN, 1BSF, 3FJ5 and 2OCT, with partial amyloid-like substructures. All of these entries are documented as amyloidogenic in the literature.}
	\end{center}
\end{figure}

\section*{Conclusions}

We have demonstrated the validity of three geometric structural selection rules, which identify amyloid fibrils and plaques in the Protein Data Bank. Additionally, these rules find non-amyloid soluble proteins, among which we have identified several amyloidogenic ones by scanning the literature. We believe that the great majority of the soluble proteins in the list show also --- mostly still undocumented --- amyloidogenic properties.

\section*{Data availability} 

The automatically updated web page is available at \url{https://pitgroup.org/amyloid/}. The page contains the list of the PDB entries found by our program, each entry is given in the graphical form, hyperlinked to the structures at the RCSB PDB site \url{https://www.rcsb.org/pdb}.

The Python source code of the software program, which generates the PDB\_Amyloid list is available at \url{http://uratim.com/amyloid/amyloid\_pit.zip}.

The page \url{https://pitgroup.org/amyloid/} contains not only the graphical representation of the proteins found but also a list of their PDB codes at 
\url{https://pitgroup.org/apps/amyloid/amyloid_list}.

\section*{Acknowledgments}
 KT and BV were partially funded by the VEKOP-2.3.2-16-2017-00014 program, supported by the European Union and the State of Hungary, co-financed by the European Regional Development Fund, and by the 
 European Union, co-financed by the European Social Fund (EFOP-3.6.3-VEKOP-16-2017-00002). VG was partially funded the NKFI-126472 grant of the National Research, Development and Innovation Office of Hungary. 
\bigskip 

\noindent Conflict of Interest: The authors declare no conflicts of interest.

\section*{Author contributions:} VG initiated the study and analyzed results. BV created the web interface and the update-mechanism. KT designed and programmed the geometric filtering algorithm and fine-tuned the geometric constraints. 



\end{document}